\documentclass[prl,twocolumn,preprintnumbers,amsmath,amssymb,nofootinbib]{revtex4}
\usepackage{subfigure}
\usepackage{graphicx}
\usepackage{bm}

\newcommand{\ee}{\begin{equation}}
 \newcommand{\eee}{\end{equation}}
\newcommand{\ea}{\begin{eqnarray}}
 \newcommand{\eea}{\end{eqnarray}}

\newcommand{\om}{\Omega_{\rm m}}
\newcommand{\od}{\Omega_{\rm d}}
\newcommand{\ob}{\Omega_{\rm b}}
\newcommand{\og}{\Omega_\gamma}

\newcommand{\olam}{\Omega_{\Lambda}}

\newcommand{\ode}{\Omega_{\rm d}^e}

\preprint{HD-THEP-06-19}
\begin{document}

\author{Michael Doran}
\email{M.Doran@thphys.uni-heidelberg.de}
\affiliation{Institut f\"ur  Theoretische Physik, Philosophenweg 16, 69120 Heidelberg, Germany}
\author{Steffen Stern}
\email{S.Stern@thphys.uni-heidelberg.de}
\affiliation{Institut f\"ur  Theoretische Physik, Philosophenweg 16, 69120 Heidelberg, Germany}
\author{Eduard Thommes}
\email{E.Thommes@thphys.uni-heidelberg.de}
\affiliation{Institut f\"ur  Theoretische Physik, Philosophenweg 16, 69120 Heidelberg, Germany}

\title{Baryon Acoustic Oscillations and Dynamical Dark Energy}

\begin{abstract}
We compute the impact of dark energy at last scattering on measurements
of baryon acoustic oscillations (BAOs). We show that an early dark energy component can contribute a systematic uncertainty to BAO measurements of up to $2.5\%$. Whilst this effect turns out to only slightly affect current BAO surveys, the results of future BAO surveys might become biased. We find that BAO surveys alone appear unable to resolve this systematic uncertainty, so supplementary measurements are necessary.
\end{abstract}

\maketitle

\section{Introduction}
Current observations 
\cite{Astier:2005qq,Riess:2004nr,Spergel:2006hy,Readhead:2004gy,Goldstein:2002gf,Tegmark:2003ud} 
suggest that roughly 70\% of our Universe consists of dark energy.  
It seems rather peculiar that in the current Universe, dark matter
and dark energy contribute roughly equally towards the total
energy density. While cosmological constant models so far fail
to address this issue, dynamical dark energy models
 \cite{Wetterich:fm,Ratra:1987rm,Caldwell:1997ii} 
might offer an explanation. 
If the fractional energy density $\od(a)$
of dark energy evolved within one or two orders of magnitude 
 similar to that of dark matter, it comes less to a surprise that 
this is also the case today.  In terms of (effective \cite{Doran:2002bc}) scalar field models,
this can e.g. be achieved using exponential potentials or 
couplings between dark energy and dark matter.  Typically,
such models contribute $\ode < 5 \%$ towards the energy
density at early times $z \gtrsim 10$ \cite{Doran:2006kp,Bean:2001wt,Linder:2006ud}. 

An important feature of such dark energy models is the change
of cosmological distances, including the sound horizon at last scattering \cite{Doran:2000jt}.
In this paper, we will therefore extend the analysis of \cite{Eisenstein:2005su}
concerning the baryon acoustic oscillations (BAO) to the case of early dark energy
models. 
The importance of BAO measurements as
a probe of dark energy has for example been pointed out in
\cite{Linder:2003ec,Seo:2003pu}. Furthermore, the BAO data extracted
from the SDSS luminous red galaxy survey \cite{Eisenstein:2005su} was used to
constrain parameterizations of dynamical dark energy models
\cite{Wang:2006ts,Dick:2006ev,Gong:2005de,Ichikawa:2005nb}, where
it supplemented Cosmic Microwave Background  (CMB) and Supernovae Ia  (Sne Ia) measurements.

\section{Baryon acoustic oscillations}
The acoustic oscillations of baryons and photons in the primeval plasma lead 
to pronounced peaks in the multipole spectrum of the CMB.
 At last scattering, the pattern of over and under densities in the
baryon fluid remains roughly speaking imprinted on scales of the sound horizon
at decoupling. In particular, baryonic oscillations leave an imprint on the matter
power spectrum. 
In \cite{Eisenstein:2005su} Eisenstein et al. report the detection of 
acoustic oscillations in the redshift-space correlation function of the SDSS
LRG sample.  Starting with a fiducial cosmology of $\om = 0.3$, $\olam = 0.7$ and $h=0.7$, they obtain
constraints on $\Omega_m h^2$ and $D_V(0.35)$ by fitting numerical simulations to the observed spectrum. $D_V(0.35)$ is the distance to $z=0.35$,
\ee
D_V(z) \equiv \left[ D_M(z)^2 \frac{cz}{H(z)} \right]^{1/3} 
\eee
which is chosen such that it correctly accounts for the Alcock-Paczynski effect \cite{Alcock:1979mp}. This effect states that the comoving angular diameter distance, $D_M(z)$, and the radial distance vary differently with cosmology. Generally speaking, a homogeneous quantity (like $D_V$) which is derived from an observed three-dimensional Galaxy distribution has to be defined such that it incorporates quantities which vary like angular and radial distance measures in the correct relative power, i.e. it has to be quadratic in angular and linear in radial distance measures. 

The two measured quantities $\Omega_m h^2$ and $D_V(0.35)$ can be combined to a single parameter ($A$) which Eisenstein et al. measure as \cite{Eisenstein:2005su} \footnote{Note that this result is only valid if we are close to the fiducial cosmology $\om = 0.3$, $\olam = 0.7$.}
\ee
 A \equiv D_V(z=0.35) \frac{\sqrt{\Omega_m H_0^2}}{0.35c} = 0.469 \pm 0.017. 
\eee
The advantage of using the $A$-parameter is its explicite dependence on $E(z)$ and $\om$ and its independence of the Hubble constant $H_0$, more precisely \cite{Eisenstein:2005su}
\ee
 A = \sqrt{\om} E(0.35)^{-1/3} \left[ \frac{1}{0.35} \int_0^{0.35} \frac{dz}{E(z)} \right]^{2/3},
\eee
where $E(z)=H(z)/H_0$. This value in combination with CMB and Sne Ia 
data was used to constrain the equation of state of dark
energy \cite{Wang:2006ts,Dick:2006ev,Gong:2005de,Ichikawa:2005nb},
alternative dark energy models
\cite{Cardone:2005aa,Cardone:2005ut,Capozziello:2005pa,Ellis:2006nd},
as well as neutrinos and extra light particle masses
\cite{Cirelli:2006kt,Kristiansen:2006ec}.

\section{Effects of Early Dark Energy}
Since the fraction of early dark energy $\ode$ is constrained by the CMB 
and Nucleosynthesis to $\ode \lesssim 5 \%$ \cite{Doran:2006kp}, it suffices to compute 
the effect of $\ode$ on $A$ to first order, i.e. 
\ee
\Delta A = \ode \frac{\partial A}{\partial \ode}_{|\ode = 0}.
\eee 
The main effect of early dark energy on the BAO measurement is a change
of the sound horizon that influences the positions of the acoustic
peaks. Assuming a constant early dark energy fraction $\ode$,
the sound horizon at last scattering is given by the analytic formula \cite{Doran:2006kp} 
\begin{multline} \label{eqn::rs}
r_s = \frac{4 \sqrt{1-\ode}}{3 H_0} \sqrt{\frac{\og^0}{\ob^0 \om^0}}\\ \times 
\ln \frac{ \sqrt{ 1 + R^{-1}_{ls}} + \sqrt{R^{-1}_{ls} + R^{-1}_{equ.}} } { 1 + \sqrt{R^{-1}_{equ.}}}.
\end{multline}
Here, $R_{ls}$ and $R_{equ.}$ is the photon to baryon ratio at last scattering and matter-radiation
equality respectively, $R =\frac{4}{3}\frac{\rho_{\gamma}}{\rho_{b}}$.
This result differs only by a factor of  $\sqrt{1-\ode}$ from that at vanishing $\ode$ \cite{Hu:1994uz}
\ee \label{eqn::rs_de}
r_s(\ode) = r_s(\ode = 0) \sqrt{1-\ode}.
\eee
The actual baryonic sound horizon measured in BAO surveys is slightly higher since baryonic waves keep on propagating until the end of the Compton drag epoch, whereas the CMB oscillations already freeze at last scattering \cite{Hu:1995en,Eisenstein:1997ik}. The resulting increase of almost $5\%$ in the BAO sound horizon is, however, independent of early dark energy and can be regarded as a further systematic effect which has to be considered in BAO surveys. 

Further note that the sound horizon scale imprinted in the baryon correlation function remains basically unchanged during structure growth \cite{Eisenstein:2005su, Seo:2005ys}. Hence, while early dark energy might shift both $r_s$ and the distance to the last scattering surface and thus might not be fully recognized in a pure CMB measurement \cite{Linder:2006uf}, BAO provides a direct measure of $r_s$ without possible cancelations, since in typical early dark energy models the late time evolution (between $z=0$ and $z=0.35$) is independent of $\ode$.

To estimate the importance of the effect of early dark energy we re-examine the BAO measurement performed in \cite{Eisenstein:2005su}. An early dark energy component will basically cause two effects, it will change the scale of the correlation-function and (rather mildly) suppress the small scale power of linear
fluctuations \cite{Caldwell:2003vp}. The latter effect mimics to some extend a running spectral index. We assume that this effect is negligible\footnote{Lacking N-body simulations for an early dark energy universe, we are yet unable to quantify the suppression on small scales.} compared to the change
in the sound horizon. An analysis of the BAO data including a running spectral index could in the future provide an easy way to estimate the magnitude of this second contribution.

Thus, to first order the only effect of early dark energy is a change in the scale of the correlation-function, and this scale is in the Eisenstein treatment \cite{Eisenstein:2005su} 1:1 related to a change in the distance measure $D_V$,
\ee \label{eqDvChange}
D_V(z=0.35, \ode) = D_V(z=0.35, \ode=0) \sqrt{1-\ode}.
\eee
In a simplified model, assuming that all galaxies in the survey are located at the same redshift, the physical explanation for that shift is quite simple.
The smaller sound horizon compared to a $\ode=0$
universe leads to a shift in the observed angle of the baryon acoustic
oscillations in the matter distribution. In order to observe the
oscillations under the same angle on the sky today, our distance to
$z=0.35$ must hence be smaller by a factor of $\sqrt{1-\ode}$ to
compensate for the shift in $r_s$.

As $\om h^2$ is independent of $\ode$, early dark energy will to first order only affect $D_V$ according to \eqref{eqDvChange}. We therefore get 
\ee \label{eqAChange}
A(\ode) = A(\ode =0) \sqrt{1-\ode},
\eee
and 
\ee
\Delta A = -\frac{1}{2} A \ode.
\eee
As $\ode \lesssim 5\%$, early dark energy will systematically reduce the BAO-results $D_V$ and $A$ by up to $2.5\%$,
\ee
\frac{| \Delta A |}{A} \approx \frac{| \Delta D_V |}{D_V} \lesssim  2.5\%.
\eee

In the case of \cite{Eisenstein:2005su}, this would mean a change of $A$ due to early dark energy 
\ee
| \Delta A | \lesssim  0.01
\eee
resulting in a slightly enhanced uncertainty in $A$,
\ee
A = 0.469 \pm 0.017 \longrightarrow A = 0.469 \pm 0.020
\eee
Hence, recent BAO galaxy surveys are just slightly affected by early dark energy.

\section{Consequences for future BAO surveys}

As also pointed out in \cite{Linder:2006uf}, early dark energy leads to a miscalibration in BAO measurements which cannot be resolved by BAO measurements alone. While the resulting systematic error of up to $2.5\%$ is small compared to an error in the range of $4-5\%$ for current BAO studies, the accuracy of future BAO surveys might reach an accuracy of $1-2\%$ \cite{Blake:2003rh,Glazebrook:2005mb,Eisenstein:2006nk,Huff:2006gs}. Without determining a possible early dark energy component by complementary measurements (CMB, nucleosynthesis), the accuracy of future BAO surveys is hence dominated by the systematic error of early dark energy. If neglected, these surveys will become biased and hence possibly lead to an incorrect determination of cosmological parameters.

\section{Determination of early dark energy with BAO surveys}

In the preceding sections we have assumed that BAO measurements are used to obtain constraints on late-time cosmological parameters. Adding the early dark energy dependence to BAO at first sight provides us with the possibility to constrain early dark energy with BAO measurements. However, running a Monte Carlo simulation with the WMAP 3-year data \cite{Spergel:2006hy}, Sne Ia \cite{Riess:2004nr} and the modified Eisenstein BAO value showed that the bounds on $\ode$ coming from the CMB are considerably stronger than those coming from BAO: adding the $\ode$ dependence according to \eqref{eqAChange} gives no difference to the results obtained in \cite{Doran:2007ep}. The picture stays the same when reducing the error of A to $1\%$, thus indicating that also future BAO surveys need to be supplemented by alternative future measurements which constrain $\ode$ with a higher precission. For example, the upcoming Planck mission might provide considerably stronger bounds on $\ode$.

\section{Conclusions}
Considering impacts of early dark energy to the matter power spectrum,
we estimated possible changes to Baryon Acoustic Peak measurements. We found that for the case of Eisenstein et al. these changes are roughly a factor
of two smaller than the experimental errors on the 
$A$-parameter measured by Eisenstein et al. However, for future BAO surveys the systematic error caused by early dark energy might dominate the results, thus demanding subsidiary determinations of $\ode$.

\section{Acknowledgements}
We thank Georg Robbers for his support with the Monte Carlo simulations.



\begin{thebibliography}{}

\bibitem{Astier:2005qq}
  P.~Astier {\it et al.},
  Astron.\ Astrophys.\  {\bf 447}, 31 (2006)
  [arXiv:astro-ph/0510447].

\bibitem{Riess:2004nr}
A.~G.~Riess {\it et al.}  [Supernova Search Team Collaboration],
Astrophys.\ J.\  {\bf 607}, 665 (2004)
[arXiv:astro-ph/0402512].

\bibitem{Spergel:2006hy}
  D.~N.~Spergel {\it et al.},
   ``Wilkinson Microwave Anisotropy Probe (WMAP) three year results:
  arXiv:astro-ph/0603449.

\bibitem{Readhead:2004gy}
A.~C.~S.~Readhead {\it et al.},
Astrophys.\ J.\  {\bf 609} (2004) 498
[arXiv:astro-ph/0402359].

\bibitem{Goldstein:2002gf}
J.~H.~Goldstein {\it et al.},
Astrophys.\ J.\  {\bf 599}, 773 (2003)
[arXiv:astro-ph/0212517].

\bibitem{Tegmark:2003ud}
M.~Tegmark {\it et al.}  [SDSS Collaboration],
Phys.\ Rev.\ D {\bf 69} (2004) 103501
[arXiv:astro-ph/0310723].

\bibitem{Wetterich:fm}
C.~Wetterich,
Nucl.\ Phys.\ B {\bf{302}}, 668  (1988)

\bibitem{Ratra:1987rm}
B.~Ratra and P.~J.~Peebles,
Phys.\ Rev.\ D {\bf{37}}, 3406  (1988)

\bibitem{Caldwell:1997ii}
R.~R.~Caldwell,~R.~Dave and P.~J.~Steinhardt,
Phys.\ Rev.\ Lett.\  {\bf{80}}, 1582 (1998)

\bibitem{Doran:2002bc}
  M.~Doran and J.~Jaeckel,
  Phys.\ Rev.\ D {\bf 66}, 043519 (2002)
  [arXiv:astro-ph/0203018].


\bibitem{Doran:2006kp}
  M.~Doran and G.~Robbers,
  JCAP {\bf 0606}, 026 (2006)
  [arXiv:astro-ph/0601544].


\bibitem{Bean:2001wt}
  R.~Bean, S.~H.~Hansen and A.~Melchiorri,
  Phys.\ Rev.\ D {\bf 64}, 103508 (2001)
  [arXiv:astro-ph/0104162].

\bibitem{Linder:2006ud}
  E.~V.~Linder,
  Astropart.\ Phys.\  {\bf 26}, 16 (2006)
  [arXiv:astro-ph/0603584].

\bibitem{Doran:2000jt}
  M.~Doran, M.~J.~Lilley, J.~Schwindt and C.~Wetterich,
  Astrophys.\ J.\  {\bf 559}, 501 (2001)
  [arXiv:astro-ph/0012139].

\bibitem{Eisenstein:2005su}
  D.~J.~Eisenstein {\it et al.},
  Astrophys.\ J.\  {\bf 633}, 560 (2005)
  [arXiv:astro-ph/0501171].

\bibitem{Linder:2003ec}
  E.~V.~Linder,
  Phys.\ Rev.\ D {\bf 68}, 083504 (2003)
  [arXiv:astro-ph/0304001].

\bibitem{Seo:2003pu}
  H.~J.~Seo and D.~J.~Eisenstein,
  Astrophys.\ J.\  {\bf 598}, 720 (2003)
  [arXiv:astro-ph/0307460].

\bibitem{Wang:2006ts}
  Y.~Wang and P.~Mukherjee,
  Astrophys.\ J.\  {\bf 650}, 1 (2006)
  [arXiv:astro-ph/0604051].

\bibitem{Dick:2006ev}
  J.~Dick, L.~Knox and M.~Chu,
  JCAP {\bf 0607}, 001 (2006)
  [arXiv:astro-ph/0603247].

\bibitem{Gong:2005de}
  Y.~g.~Gong and Y.~Z.~Zhang,
  Phys.\ Rev.\ D {\bf 72}, 043518 (2005)
  [arXiv:astro-ph/0502262].

\bibitem{Ichikawa:2005nb}
  K.~Ichikawa and T.~Takahashi,
  Phys.\ Rev.\ D {\bf 73}, 083526 (2006)
  [arXiv:astro-ph/0511821].

\bibitem{Alcock:1979mp}
  C.~Alcock and B.~Paczynski,
  Nature {\bf 281}, 358:359 (1979).

\bibitem{Cardone:2005aa}
  V.~F.~Cardone, A.~Troisi and S.~Capozziello,
  Phys.\ Rev.\ D {\bf 72}, 043501 (2005)
  [arXiv:astro-ph/0506371].


\bibitem{Capozziello:2005pa}
  S.~Capozziello, V.~F.~Cardone, E.~Elizalde, S.~Nojiri and S.~D.~Odintsov,
  Phys.\ Rev.\ D {\bf 73}, 043512 (2006)
  [arXiv:astro-ph/0508350].

\bibitem{Cardone:2005ut}
  V.~F.~Cardone, C.~Tortora, A.~Troisi and S.~Capozziello,
  Phys.\ Rev.\ D {\bf 73}, 043508 (2006)
  [arXiv:astro-ph/0511528].

\bibitem{Ellis:2006nd}
  J.~R.~Ellis, N.~E.~Mavromatos, V.~A.~Mitsou and D.~V.~Nanopoulos,
  Astropart.\ Phys.\  {\bf 27}, 185 (2007)
  [arXiv:astro-ph/0604272].

\bibitem{Cirelli:2006kt}
  M.~Cirelli and A.~Strumia,
  JCAP {\bf 0612}, 013 (2006)
  [arXiv:astro-ph/0607086].

\bibitem{Kristiansen:2006ec}
  J.~R.~Kristiansen, H.~K.~Eriksen and O.~Elgaroy,
  arXiv:astro-ph/0608017.

\bibitem{Hu:1994uz}
  W.~Hu and N.~Sugiyama,
  Astrophys.\ J.\  {\bf 444} (1995) 489
  [arXiv:astro-ph/9407093].

\bibitem{Hu:1995en}
  W.~Hu and N.~Sugiyama,
  Astrophys.\ J.\  {\bf 471}, 542 (1996)
  [arXiv:astro-ph/9510117].

\bibitem{Eisenstein:1997ik}
  D.~J.~Eisenstein and W.~Hu,
  Astrophys.\ J.\  {\bf 496}, 605 (1998)
  [arXiv:astro-ph/9709112].

\bibitem{Seo:2005ys}
  H.~J.~Seo and D.~J.~Eisenstein,
  Astrophys.\ J.\  {\bf 633}, 575 (2005)
  [arXiv:astro-ph/0507338].

\bibitem{Linder:2006uf}
  E.~V.~Linder,
  arXiv:astro-ph/0610173.

\bibitem{Caldwell:2003vp}
  R.~R.~Caldwell, M.~Doran, C.~M.~Mueller, G.~Schaefer and C.~Wetterich,
  Astrophys.\ J.\  {\bf 591}, L75 (2003)
  [arXiv:astro-ph/0302505].

\bibitem{Blake:2003rh}
  C.~Blake and K.~Glazebrook,
  Astrophys.\ J.\  {\bf 594}, 665 (2003)
  [arXiv:astro-ph/0301632].

\bibitem{Glazebrook:2005mb}
  K.~Glazebrook and C.~Blake,
  Astrophys.\ J.\  {\bf 631}, 1 (2005)
  [arXiv:astro-ph/0505608].

\bibitem{Eisenstein:2006nk}
  D.~J.~Eisenstein, H.~j.~Seo, E.~Sirko and D.~Spergel,
  arXiv:astro-ph/0604362.

\bibitem{Huff:2006gs}
  E.~Huff, A.~E.~Schulz, M.~White, D.~J.~Schlegel and M.~S.~Warren,
  Astropart.\ Phys.\  {\bf 26}, 351 (2007)
  [arXiv:astro-ph/0607061].

\bibitem{Doran:2007ep}
  M.~Doran, G.~Robbers and C.~Wetterich,
  Phys.\ Rev.\  D {\bf 75}, 023003 (2007)
  [arXiv:astro-ph/0609814].

\end{thebibliography}
\end{document}